\documentclass{gNST2e}

\usepackage{amsmath,amssymb}
\usepackage{amsthm}
\usepackage{bigints}
\usepackage[skip=0pt]{caption}
\usepackage{color}
\usepackage{epstopdf}  
\usepackage{enumerate,epsf,etoolbox}
\usepackage{graphicx}
\usepackage{lmodern}
\usepackage{mathtools}
\usepackage{multirow}
\usepackage{psfrag}
\usepackage{subfigure}  
\usepackage{url}  
\usepackage[table]{xcolor}  
\usepackage{xr}

\theoremstyle{definition}

\newtheorem*{pf}{Proof}
\newtheorem{thm}{Theorem}[section]

\theoremstyle{definition}
\newtheorem{defn}{Definition}[section]

\newcommand{\cF}{\mathcal{F}}
\newcommand{\cN}{\mathcal{N}}

\newcommand{\cU}{\mathcal{U}}
\newcommand{\mR}{\mathbb{R}}
\newcommand{\eps}{\epsilon}

\newcommand{\bed}{\begin{itemize}}
\newcommand{\eed}{\end{itemize}}
\newcommand{\ben}{\begin{enumerate}}
\newcommand{\een}{\end{enumerate}}

\renewcommand{\qedsymbol}{\rule{0.7em}{0.7em}}

\DeclareMathOperator*{\argmax}{arg\,max}
\DeclareMathOperator*{\argmin}{arg\,min}
\DeclareMathOperator*{\converge}{\rightarrow}
\DeclareMathOperator*{\multsums}{\sum}

\begin{document}


\title{Variable screening based on Gaussian Centered L-moments}

\author{
\name{Hyowon An\textsuperscript{a}$^{\ast}$\thanks{$^\ast$Corresponding author. Email: ahwbest@gmail.com},
Kai Zhang\textsuperscript{a},
Hannu Oja\textsuperscript{b}
and J. S. Marron\textsuperscript{a}}
\affil{
\textsuperscript{a}The University of North Carolina at Chapel Hill, Chapel Hill, North Carolina, USA;\\
\textsuperscript{b}University of Turku, 20500 Turku, FI}
\received{v1.0 released March 2018}
}

\maketitle

\begin{abstract}
An important challenge in big data is identification of important variables. In this paper, we propose methods of discovering variables with non-standard univariate marginal distributions. The conventional moments-based summary statistics can be well-adopted for that purpose, but their sensitivity to outliers can lead to selection based on a few outliers rather than distributional shape such as bimodality. To address this type of non-robustness, we consider the L-moments. Using these in practice, however, has a limitation because they do not take zero values at the Gaussian distributions to which the shape of a marginal distribution is most naturally compared. As a remedy, we propose Gaussian Centered L-moments which share advantages of the L-moments but have zeros at the Gaussian distributions. The strength of Gaussian Centered L-moments over other conventional moments is shown in theoretical and practical aspects such as their performances in screening important genes in cancer genetics data.
\end{abstract}


\begin{classcode}
62G30; 62G35;
\end{classcode}

\section{Introduction}\label{sec:introduction}

\begin{figure}
\begin{center}
\begin{tabular}{c}
\includegraphics[width=0.96\textwidth]{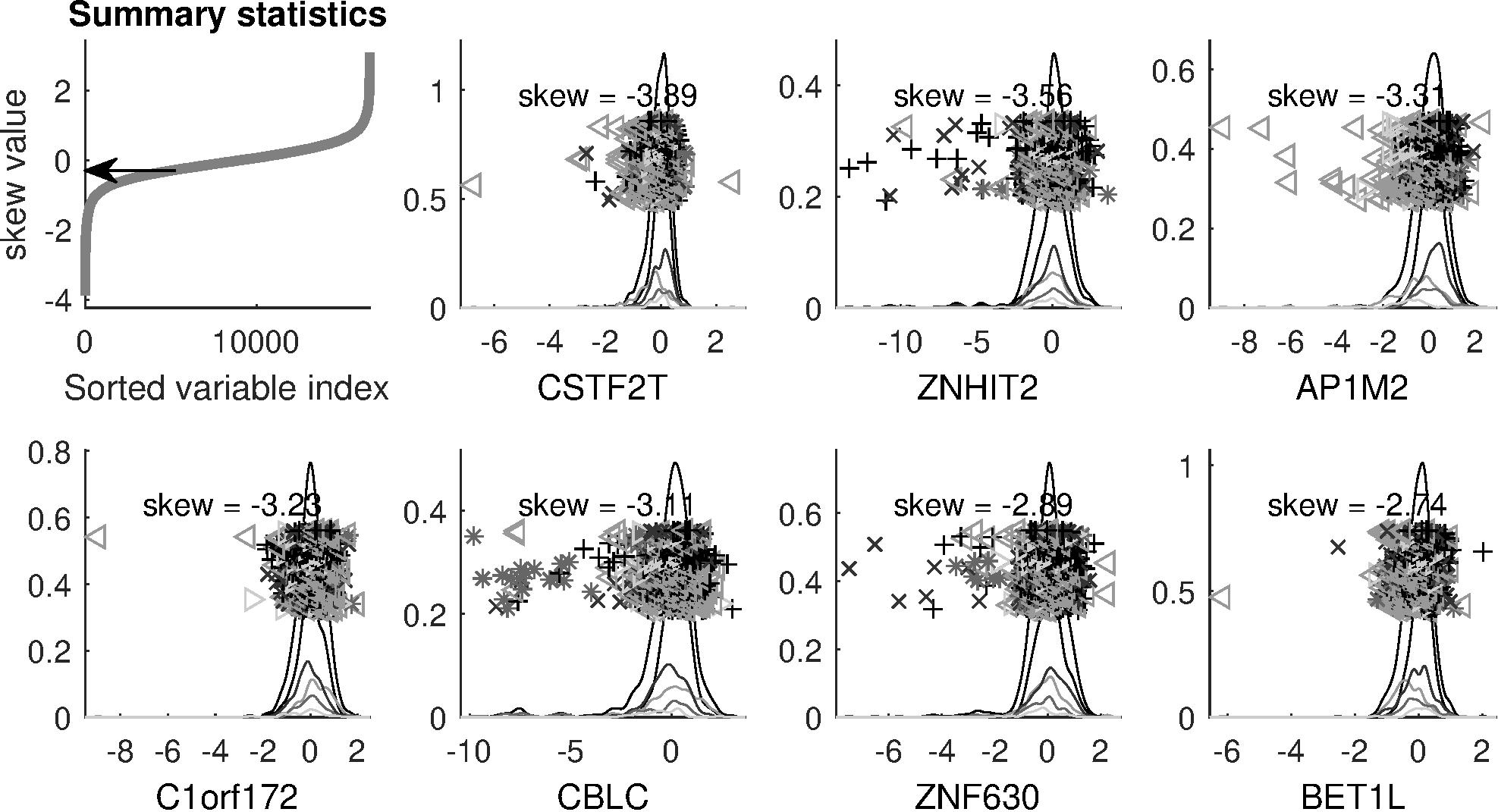}\\\\
\includegraphics[width=0.96\textwidth]{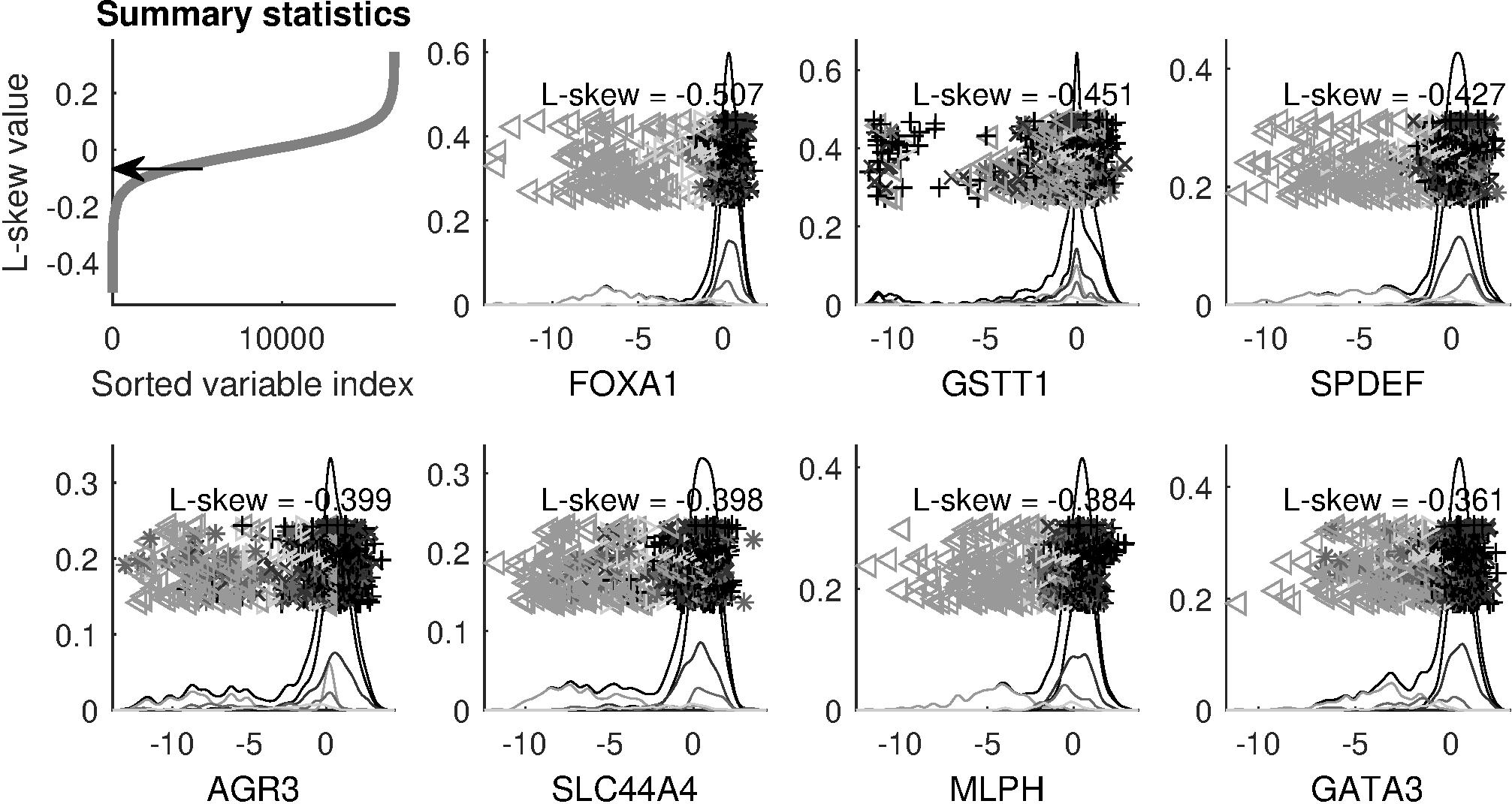}
\end{tabular}
\end{center}
\caption{The marginal distribution plots of seven genes, i.e. variables, with the smallest conventional sample skewness (top two rows) and L-skewness values (bottom two rows). Each of the seven genes in the top two rows is driven by a few strong outliers that tend to obscure distributional shapes, while the seven genes in the bottom two rows appear because of distributional structure.}
\label{fig:mdp_bw_skew_L_skew_bottom}
\end{figure}

Data quality is an issue that is currently not receiving as much attention as it deserves in the age of big data and data science. Traditional careful analysis of small data sets involves a study of marginal distributions, which easily finds data quality challenges such as skewness and suggests remedies such as the Box-Cox transformation \citep{Box1964}. Direct implementation of this type of operation is challenging with high dimensional data, as there are too many marginal distributions to individually visualize. Automatic methods such as those of \citet{Feng2016} are available, but direct visualization is still very useful. This hurdle can be overcome by using summary statistics to select a representative set for visualization and potential remediation. Conventional summaries such as the sample mean, standard deviation, skewness and kurtosis can be very useful for this process. However, as seen in Figure \ref{fig:mdp_bw_skew_L_skew_bottom} and Section \ref{sec:TCGA_data}, those have some limitations for this purpose, e.g. they can be strongly influenced by outliers. In some situations outliers are important and well worth finding. But in other cases summaries that are dominated by outliers, e.g. the conventional summaries stated earlier, can miss more important distributional features of variables such as skewness and bimodality. Such variables can be of keen interest in cancer research.

As measures of skewness and bimodality of a distribution, we first introduce the conventional theoretical (moment-based) \textit{skewness} $\gamma_1$ and (excess) \textit{kurtosis} $\gamma_2$ of a random variable $X$ defined as
\begin{equation}\label{eqn:convention_skewness_kurtosis}
\gamma_1=\frac{E\left(X-EX\right)^3}{\left\{E\left(X-EX\right)^2\right\}^{3/2}},\quad
\gamma_2=\frac{E\left(X-EX\right)^4}{\left\{E\left(X-EX\right)^2\right\}^2}-3.
\end{equation}
See \citet{Pearson1905} for the origin of the word \textit{excess}. We use the word \textit{conventional} to distinguish these traditional measures from other measures that will be introduced in upcoming sections. The conventional \textit{sample skewness} $\hat{\gamma}_1$ and \textit{kurtosis} $\hat{\gamma}_2$ are typical sample-moment-based estimators of the conventional theoretical skewness and kurtosis, and are defined as
\begin{equation}\label{eqn:sample_skewness_kurtosis}
\hat{\gamma}_1=\frac{\frac{1}{n}\sum_{i=1}^n\left(X_i-\bar{X}\right)^3}
{\left(\frac{1}{n}\sum_{i=1}^n\left(X_i-\bar{X}\right)^2\right)^{3/2}},
\quad\hat{\gamma}_2=\frac{\frac{1}{n}\sum_{i=1}^n\left(X_i-\bar{X}\right)^4}
{\left(\frac{1}{n}\sum_{i=1}^n\left(X_i-\bar{X}\right)^2\right)^2}-3.
\end{equation}
Many papers including \citet{Brys2004,Brys2006} pointed out that those estimators can be highly affected by outliers in real data analysis.

In Section \ref{sec:TCGA_data}, the limitation of conventional summary statistics in practice is demonstrated in detail using a modern high dimensional data set from cancer research. These data are part of the TCGA project \citep{Weinstein2013}, and were first studied in \citet{Ciriello2015} and \citet{Hu2015}. The precise version of the data here was used in \citet{Feng2016}. The data include gene expression profiles of 16,615 genes and 817 breast cancer patients, each of whom is classified according to five cancer subtypes of major importance in modern cancer treatment. While much is known about this data, as discussed in \citet{Feng2016}, the sheer data size means that there have only been relatively cursory studies of the marginal, or individual gene, distributions. In this study we do a much deeper search for genes with unexpected marginal structure, especially those related to non-Gaussianity. This can yield relationships between genes and biologically meaningful features such as breast cancer subtypes.

The top two rows of Figure \ref{fig:mdp_bw_skew_L_skew_bottom} show the marginal distributions of seven variables, i.e. genes, with the smallest conventional sample skewness values. The upper left plot shows the sample quantile curve of these summary statistic values as a function of their ranks. The left arrow in the plot indicates that the seven shown marginal distribution plots have the smallest sample skewness values. These remaining plots are sorted in ascending order of the sample skewness values that are given near the top of each plot. Each symbol with a different shade of gray represents a breast cancer patient by subtypes; see Table \ref{tab:breast_cancer_subtype}. The height of each symbol provides visual separation based on the order of observations in the data set. The black solid lines are kernel density estimates of marginal distributions and gray solid lines are sub-densities corresponding to different subtypes. The aim of including subtype-related information in marginal distribution plots is to check whether skewness or bimodality of a marginal distribution comes from different distributions of cancer subtypes. Some genes can have skewed distributions that do not result from cancer subtypes, and we check whether such genes indeed convey biological meanings in a quantitative way in Section \ref{subsec:GSEA}.

Figure \ref{fig:mdp_bw_skew_L_skew_bottom} shows that even though the genes with the smallest sample skewness values were selected, genes such as CSTF2T, C1orf172 and BET1L have a couple of outliers on their left sides rather than distributional skewness to the left. Only the gene CBLC has several Her2-type samples forming a small cluster on the left side, but its marginal distribution seems to be symmetric rather than skewed. The sample skewness seems inadequate for effectively screening interesting genes in terms of skewness of their distributional bodies. The color version of Figure \ref{fig:mdp_bw_skew_L_skew_bottom} is reproduced in Section \ref{supp:sec:TCGA_supplement} of the Supplementary Material.

\begin{table}[!b]
\begin{center}
\begin{tabular}{c||c|c|c|c|c}
Subtype&LumA&LumB&Her2&Basal&Normal-like\\\hline
Symbol&{\color[rgb]{0,0,0}+}&{\color[rgb]{0.2,0.2,0.2}$\times$}&{\color[rgb]{0.4,0.4,0.4}*}&
{\color[rgb]{0.6,0.6,0.6}$\vartriangleleft$}&{\color[rgb]{0.8,0.8,0.8}$\vartriangleright$}
\end{tabular}
\end{center}
\caption{The symbols and gray-levels corresponding to the 5 breast cancer subtypes in the marginal distribution plots.}
\label{tab:breast_cancer_subtype}
\end{table}

This challenge is well addressed using another notion of skewness as shown in the bottom two rows of Figure \ref{fig:mdp_bw_skew_L_skew_bottom}. That alternative robust measure of skewness is related to the \textit{L-moments} whose $r$-th term is defined in \citet{Hosking1990} as
\begin{equation}\label{eqn:L_moments_defn_1}
\lambda_r=\frac{1}{r}\sum_{k=0}^{r-1}(-1)^k\left(\!
\begin{array}{c}r-1\\k\end{array}\!\right)E\!\left(X_{(r-k):r}\right)
\end{equation}
where $X_{i:n}$ is the $i$-th order statistic of the random sample $X_1,X_2,\cdots,X_n$. We call these the \textit{classical L-moments} to distinguish them from other moments that we will study in Section \ref{sec:GCL_moments}. The \textit{L-skewness} $\lambda_3^*$ \citep{Hosking1990} is defined as the ratio of the third and second L-moments, $\lambda_3^*=\lambda_3/\lambda_2$. The L-moments are known to have some robustness properties against outliers, and this is confirmed in the bottom two rows of Figure \ref{fig:mdp_bw_skew_L_skew_bottom}. The seven genes with the smallest \textit{sample L-skewness} \citep{Hosking1990} values show distributional skewness in contrast to the genes screened by the conventional skewness. Especially, all the genes except GSTT1 have clusters of Basal-type samples (gray triangles) and the gene GSTT1 has a small cluster of LumA samples (black crosses). These show that the L-skewness screens genes with distributional skewness especially coming from cancer subtypes rather than those with outliers, which is desirable in the sense that the former is clinically much more interesting than the latter.

With the goal of screening interesting non-Gaussian variables, a limitation of the classical L-moments is that they are not centered at the Gaussian distributions, which typically and frequently appear in real data. Instead the classical L-moments are zero at the uniform distributions, which hinders interpreting the signs and magnitude of the L-moments especially in terms of the critical notion of kurtosis represented by the \textit{L-kurtosis} $\lambda_4^*=\lambda_4/\lambda_2$ \citep{Hosking1990}. The paper \citet{Hosking1990} proposed using the L-skewness to perform the goodness-of-fit test for Gaussianity against skewed distributions, but it did not consider using the L-kurtosis against kurtotic alternative distributions. A simple approach is to subtract the L-kurtosis value at the Gaussian distributions from itself. However, this can result in loss of theoretical soundness of the definition of the L-moments because they are no longer differences of expected order statistics as in Equation (\ref{eqn:L_moments_defn_1}). Note that the conventional excess kurtosis bases its theoretical justification in the \textit{cumulant} theory (see \citet{Feller1968} and \citet{Marcinkiewicz1939}), which is more convincing than simply subtracting its value at the Gaussian distributions.

These ideas motivate development of improved Gaussian centered versions of the L-moments. The paper \citet{Decurninge2014} discovered a variant of the classical L-moments that have strength in multivariate analysis. The new moments were called \textit{Hermite L-moments} (\textit{HL-moments}). While it was seen that the HL-moments have advantages over the classical L-moments in a multivariate setting, no attention was paid to the potential of the HL-moments for univariate summary statistics, nor was its distributional centering at the Gaussian distributions mentioned. In this paper, we comprehensively investigate the possibilities of the HL-moments as univariate summaries, especially focusing on their third and fourth moments.

To this end, we propose two classes of moments that share robustness of the L-moments while also having zeros at the Gaussian distributions by their theoretical nature. The first is based on the family of Hermite polynomials, and the second uses spacing between the expected order statistics of the standard Gaussian distribution. These \textit{Gaussian Centered L-moments} are developed in Section \ref{sec:GCL_moments}. Their abilities to screen variables with interesting marginal distributions are visualized and quantitatively analyzed in Section \ref{sec:TCGA_data}. Their theoretical properties such as asymptotic Gaussianity and robustness are shown in Sections \ref{sec:estimation_GCL_moments} and \ref{sec:robustness}, respectively.

\section{Mathematical preliminaries}\label{sec:math}

Assume that two random variables $X$ and $Y$ follow cumulative distribution functions $F$ and $G$ with probability density functions $f$ and $g$, respectively. We denote the distribution of $aX+b$ for $a\neq0$ and $b\in\mR$ by $F_{a,b}$. Also, a random sample $X_1,X_2,\cdots,X_n$ is assumed to be generated from $F$, and $X_{i:n}$ denotes the i-th order statistic of the random sample, i.e. $X_{1:n}\leq X_{2:n}\leq\cdots\leq X_{n:n}$. In this paper we consider only an absolutely continuous and strictly increasing cumulative distribution function, in the sense that it is strictly increasing on its support $S(F)=\overline{\{x|0<F(x)<1\}}$ where $\overline{A}$ indicates the closure of a set $A\subset\mR$. Let $\cF$ be the class of such distribution functions and the quantile function $F^{-1}:(0,1)\rightarrow\mR$ be the inverse of $F\in\cF$. It is always assumed that the composition of $G^{-1}$ and $F$, $G^{-1}\circ F$, is defined on $S_F$. When $F$ is symmetric, we denote its \textit{point of symmetry} \citep{Doksum1975} by $m(F)$.

Various kinds of orthogonal polynomials are used throughout this paper. One is the \textit{shifted Legendre polynomials} $\left\{P_r^*|r=1,2,\cdots\right\}$ which have been comprehensively investigated in Chapter 4 of \citet{Szego1959}. The shifted Legendre polynomials are orthogonal to each other on the unit interval $(0,1)$ with respect to the weight function $w(x)=1$. Other orthogonal polynomials of interest are the \textit{Hermite polynomials} presented in Chapter 5 of \citet{Szego1959}. There are two versions of the Hermite polynomials, and here we focus on the \textit{probabilists' Hermite polynomials} $\left\{H_r|r=1,2,\cdots\right\}$ which are orthogonal to each other on the real line $\mR$ with respect to the weight function $w(x)=e^{-x^2/2}$. The first four shifted Legendre and Hermite polynomials are given in Table \ref{tab:first_four_shifted_Legendre_Hermite}.

\begin{table}
\begin{center}
\begin{tabular}{r|r}
Shifted Legendre $P_r^*:(0,1)\rightarrow\mR$&
Hermite $H_r:\mR\rightarrow\mR$\\\hline\hline
$P_0^*(u)=1$&$H_0(x)=1$\\
$P_1^*(u)=2u-1$&$H_1(x)=x$\\
$P_2^*(u)=6u^2-6u+1$&$H_2(x)=x^2-1$\\
$P_3^*(u)=20u^3-30u^2+12u-1$&$H_3(x)=x^3-3x$
\end{tabular}
\end{center}
\caption{The first four shifted Legendre and Hermite polynomials.}
\label{tab:first_four_shifted_Legendre_Hermite}
\end{table}

\subsection{L-statistics and L-moments}\label{subsec:L_statistic_L_moment}

The term \textit{L-statistic} is used to indicate a statistic in the form of a \textit{linear combination of order statistics}.  An L-statistic is generally expressed as
\begin{equation}\label{eqn:L_statistic}
\hat{\theta}_n=\frac{1}{n}\sum_{i=1}^nc_{ni}X_{i:n}
\end{equation}
where $c_{ni}$ is a function of the sample size $n$ and the rank $i$ of the order statistic $X_{i:n}$. The L-statistics were first proposed in the general research area of robust statistics; see \citet{Stigler1973} for a description of the origins of the L-statistics. Also, Section 11.4 of \citet{David2003} surveys the literature on various sets of conditions on the coefficients $\left\{c_{ni}|n\geq1,1\leq i\leq n\right\}$ and the distribution function $F$ which ensure that $\hat{\theta}_n$ almost surely converges in the limit as $n\rightarrow\infty$ to the quantity
\begin{equation}\label{eqn:L_functional}
\theta(F)=\int_{-\infty}^\infty\!xf(x)J(F(x))\,\mathrm{d}x
\end{equation}
where $J:(0,1)\rightarrow\mR$ is a measurable function, and further follow an asymptotic Gaussian distribution. We call a functional in the form of Equation (\ref{eqn:L_functional}) an \textit{L-functional}, the term used in the papers \citet{Welsh1990} and \citet{Necir2010}.

A connection between L-statistics and location, scale, skewness and kurtosis of a distribution has been made by \citet{Hosking1990}. In addition to the intuitive definition of the $r$-th L-moment in Equation (\ref{eqn:L_moments_defn_1}), the other form is given in that paper as
\begin{equation}\label{eqn:L_moments_integ}
\lambda_r=\int_{-\infty}^\infty xf(x)P_{r-1}^*(F(x))\!\,\mathrm{d}x=\int_0^1F^{-1}(u)P_{r-1}^*(u)\!\,\mathrm{d}u,
\end{equation}
which shows that the L-moments are L-functionals. \citet{Hosking1990} adopted the U-statistics-based estimators
\begin{equation}\label{eqn:sample_L_moment_U_statistic}
\hat{\lambda}_{n,r}=\left(\!\begin{array}{c}n\\r\end{array}\!\right)^{-1}
\multsums_{1\leq i_1<i_2<\cdots<i_r\leq n}\frac{1}{r}\sum_{k=0}^{r-1}(-1)^k
\left(\!\begin{array}{c}r-1\\k\end{array}\!\right)X_{i_{r-k}:n},
\end{equation}
and called these the \textit{sample L-moments}. The \textit{sample L-moment ratios} are defined as $\hat{\lambda}_{n,r}^*=\hat{\lambda}_{n,r}/\hat{\lambda}_{n,2}$ accordingly.

\subsection{Oja's criteria}\label{subsec:Oja_criteria}

When defining new measures of location, scale, skewness and kurtosis, a challenge is to ensure that the new measures reflect the intuitive meaning of those distributional properties. This challenge can be addressed by the framework of \citet{Oja1981} using stochastic dominance ideas, which are applied here. To do this we say that a function $f:I\rightarrow\mR$ is \textit{convex of order k} if $f^{(k)}(x)\geq0$ for all $x\in I$ where $I$ is an open interval and $f^{(k)}$ is the $k$-th order derivative of $f$. Note that if $X\sim F$, then $G^{-1}\circ F(X)\sim G$, so $G^{-1}\circ F$ is a natural link between the distributions $F$ and $G$.
\begin{defn}[\citet{Oja1981}]\label{defn:Oja_criteria}
The functional $\theta:\cF\rightarrow\mR$ is a
\bed
\item[a.] \textit{measure of location} in $\cF$ if $\theta\!\left(F_{a,b}\right)=a\theta(F)+b$ for all $a,b\in\mR,F\in\cF$ and $\theta(F)\leq\theta(G)$ whenever $G^{-1}\circ F$ is convex of order 0.
\item[b.] \textit{measure of scale} in $\cF$ if $\theta\!\left(F_{a,b}\right)=|a|\theta(F)$ for all $a,b\in\mR,F\in\cF$ and $\theta(F)\leq\theta(G)$ whenever $G^{-1}\circ F$ is convex of order 1.
\item[c.] \textit{measure of skewness} in $\cF$ if $\theta\!\left(F_{a,b}\right)=\text{sign}(a)\theta(F)$ for all $ a\neq0,b\in\mR,F\in\cF$ and $\theta(F)\leq\theta(G)$ whenever $G^{-1}\circ F$ is convex (of order 2). In this case, we say $F$ \textit{is not more skew to the right than} $G$.
\item[d.] \textit{measure of kurtosis} in a family of symmetric distributions $\cF_s\subset\cF$ if $\theta\!\left(F_{a,b}\right)=\theta(F)$ for all $a\neq0,b\in\mR,F\in\cF_s$ and $\theta(F)\leq\theta(G)$ whenever $F,G\in\cF_s$, $G^{-1}\circ F$ is concave on $\{x|x\leq m(F)\}$ and convex on $\{x|x>m(F)\}$. In this case, we say $F$ \textit{does not have more kurtosis than} $G$.\hfill\qedsymbol
\eed
\end{defn}

\section{Gaussian Centered L-moments}\label{sec:GCL_moments}

As mentioned in Section \ref{sec:introduction}, investigation about distributional shape is often performed relative to the Gaussian distributions. Since many distributions are aggregations of small, independent errors, they tend to have a Gaussian shape by the Central Limit Theorem. For non-Gaussian distributions such as the marginal distributions in the bottom two rows of Figure \ref{fig:mdp_bw_skew_L_skew_bottom}, it is often not easy to find a suitable transformation that yields approximately Gaussian distributions. In addition, transforming data can result in loss of some useful information such as subtype-driven bimodality. For example, the cluster of the Her2-enriched samples on the left side of the marginal distribution of the gene CBLC in Figure \ref{fig:mdp_bw_skew_L_skew_bottom} can disappear after transformation. This suggests that a search for non-Gaussianity is an important subject of exploratory data analysis. The term excess kurtosis is an example of the importance of measuring the difference between the shape of a distribution and the Gaussian distribution. Based on the zero of the excess kurtosis, distributions are often classified into \textit{platykurtic} and \textit{leptokurtic} distributions depending on their relationship with the Gaussian distributions.

However, as discussed in \citet{Hosking1990}, the L-moments satisfy
\begin{equation}\label{eqn:L_moments_uniform}
\lambda_r(\cU(a,b))=\int_0^1\!\{(b-a)x+a\}P_{r-1}^*(x)\,\mathrm{d}x=0
\end{equation}
for all $-\infty<a<b<\infty$ and $r=3,4,\cdots$. Thus the signs and absolute values of the L-moments measure the directions and magnitude of departure from the uniform distributions, which is a limitation of the L-moments as distributional summaries. This motivates introducing a definition.
\begin{defn}\label{defn:distr_center}
A sequence of functionals $\left\{\theta_r|r=1,2,\cdots\right\}$ is \textit{centered at the family of distributions} $\cF$ when it satisfies $\theta_r(F)=0$ for all $r=3,4,\cdots$ and $F\in\cF$.\hfill\qedsymbol
\end{defn}
Important functionals centered at the Gaussian distributions are the cumulants. The main goal of this research is to develop different types of moments with their distributional centers at the Gaussian family. For that purpose, we introduce the following definition.
\begin{defn}\label{defn:GCL_moments}
We call functionals $\{\theta_r:\cF\rightarrow\mR|r=1,2,\cdots\}$ \textit{Gaussian Centered L-moments}  in $\cF$ if they are L-functionals and centered at the Gaussian distributions.\hfill\qedsymbol
\end{defn}
The letter `L' generally referred to the linear combination of expected order statistics, but here it refers to any L-functionals in the form of Equation (\ref{eqn:L_functional}).

\subsection{Hermite L-moments}\label{subsec:HL_moments}

As pointed out in Equation (\ref{eqn:L_moments_uniform}), the L-moments are centered at the uniform distributions due to the orthogonality property of the shifted Legendre polynomials. This motivates us to consider adopting another family of orthogonal polynomials to locate the center of L-functionals at the Gaussian distributions. Given the $r$-th degree polynomial $J_r:(0,1)\rightarrow\mR$, define the L-functional
\[\theta_r(F)=\int_{-\infty}^\infty\!xf(x)J_{r-1}(F(x))\,\mathrm{d}x.\]
In order to be a Gaussian Centered L-moment, $\theta_r$ should satisfy
\[\theta_r(\Phi)=\int_{-\infty}^\infty\!x\phi(x)J_{r-1}(\Phi(x))\,\mathrm{d}x=0,\]
where $\Phi$ is the cumulative distribution function of the standard Gaussian distribution. This results in one possible solution
\begin{equation}\label{eqn:HL_moment}
\eta_r=\int_{-\infty}^\infty\!xf(x)H_{r-1}\left(\Phi^{-1}(F(x))\right)\,\mathrm{d}x
=\int_0^1\!F^{-1}(u)H_{r-1}\left(\Phi^{-1}(u)\right)\,\mathrm{d}u.
\end{equation}
Note that
\[\eta_r(\Phi(\cdot|\mu,\sigma^2))=\int_{-\infty}^\infty\!(\mu+\sigma x)\phi(x)H_{r-1}(x)\,\mathrm{d}x=0\]
where $\phi$ is the probability density function of the standard Gaussian distribution for all $\mu\in\mR,\sigma>0$ and $r=3,4,\cdots$ by the orthogonality of the Hermite polynomials. As mentioned in Section \ref{sec:introduction}, the paper \citet{Decurninge2014} called these L-functionals the \textit{Hermite L-moments} (\textit{HL-moments}).

Recall from Definition \ref{defn:Oja_criteria}.c and d that a measure of skewness or kurtosis should be invariant under linear transformation of a random variable. This need for scale invariance motivates us to introduce \textit{Hermite L-moment ratios} (\textit{HL-moment ratios}) defined as $\eta_r^*=\eta_r/\eta_2$ for $r=3,4,\cdots$. The \textit{HL-skewness} and \textit{HL-kurtosis} are defined as $\eta_3^*$ and $\eta_4^*$, respectively. A central issue is whether or not the HL-skewness and kurtosis actually measure the skewness and kurtosis of a distribution in the sense of Definition \ref{defn:Oja_criteria}.
\begin{thm}\label{thm:Oja_criteria_HL}
The HL-moments-based measures $\eta_1,\eta_2,\eta_3^*$ and $\eta_4^*$ satisfy Oja's criteria for measures of location, scale, skewness and kurtosis, respectively.
\end{thm}
\begin{pf}
See Subsection \ref{supp:subsec:proof_Oja_criteria_HL} of the Supplementary Material.\hfill\qedsymbol
\end{pf}
\noindent This theorem shows that the HL-moments are not only useful in a multivariate setting but also can be useful summary statistics in data analysis.

Note that the HL-moments are closely related to the inverse Edgeworth expansion
\begin{equation}\label{eqn:inverse_Edgeworth}
F^{-1}(u)\approx\mu+\sigma\Phi^{-1}(u)+\sigma\sum_{r=3}^\infty\frac{EH_r(Y)}{r!}H_{r-1}\!\left(\Phi^{-1}(u)\right)
\end{equation}
where $\mu$ and $\sigma$ are the mean and standard deviation of the random variable $X\sim F$ and $Y=(X-\mu)/\sigma$ is a standardized random variable; see \citet{Hall1983} for detailed discussion. Note that the terms $H_{r-1}\!\left(\Phi^{-1}(u)\right)$ in this expansion also appear in the definition of the HL-moments in Equation (\ref{eqn:HL_moment}). \citet{Brown1996} focused on the fact that the coefficients $EH_r(Y)$ before terms $H_{r-1}\!\left(\Phi^{-1}(u)\right)$ will be zero for the Gaussian distributions, and claimed that those coefficients can be indicators of departure from Gaussianity. Instead of directly estimating the coefficients $EH_r(Y)$, which are functions of high-order moments, they considered the least squares estimation problem
\begin{equation}\label{eqn:inverse_Edgeworth_minimization}
\left\{\tilde{\eta}_{n,r}\right\}_{r\geq1}=\argmin_{\eta_r,r\geq1}
\bigintsss_0^1\!\left\{F_n^{-1}(u)-\eta_1-\eta_2\Phi^{-1}(u)+\eta_2\sum_{r=3}^\infty\frac{\eta_r}{r!}H_{r-1}\!\left(\Phi^{-1}(u)\right)\right\}^2
\,\mathrm{d}u,
\end{equation}
where $F_n(x)=\frac{1}{n}\sum_{i=1}^nI\left(X_i\leq x\right)$ is the \textit{empirical distribution function} (\textit{EDF}), to obtain
\begin{equation}\label{eqn:sample_HL_moments_BH}
\tilde{\eta}_{n,r}=\int_0^1\!F_n^{-1}(u)H_{r-1}\left(\Phi^{-1}(u)\right)\,\mathrm{d}u
=\frac{1}{n}\sum_{i=1}^n\left(\int_{(i-1)/n}^{i/n}\!H_{r-1}\left(\Phi^{-1}(u)\right)\,\mathrm{d}u\right)X_{i:n}.
\end{equation}
Note that this estimator coincides with the estimator of Equation (7.18) in \citet{Decurninge2014} when $d=1$ in that paper. Replacing $F_n$ by $F$ in Equation (\ref{eqn:sample_HL_moments_BH}) yields the theoretical HL-moments $\eta_r$.

Based on simulation, \citet{Brown1996} claimed that $\tilde{\eta}_{n,3}/\tilde{\eta}_{n,2}$ and $\tilde{\eta}_{n,4}/\tilde{\eta}_{n,2}$ can be used as estimators of skewness and heavy-tailedness, but did not provide theoretical analysis such as verifying Oja's criteria. Particularly, they did not show that negative values of $\eta_4^*$ correspond to flat shoulders or bimodality of a distribution. Our Theorem \ref{thm:Oja_criteria_HL} shows that the HL-moment ratios $\eta_3^*$ and $\eta_4^*$ are actually measures of skewness and kurtosis in Oja's sense. In Section \ref{sec:estimation_GCL_moments}, we will present different sample estimators from $\tilde{\eta}_{n,r}$ that show better performances in our TCGA data analysis. To strengthen value of the HL-moments, we analyze the robustness properties of $\eta_3^*$ and $\eta_4^*$ in Section \ref{sec:robustness}.

\subsection{Rescaled L-moments}\label{subsec:RL_moments}

\begin{figure}[!b]
\begin{center}
\includegraphics[width=\textwidth]{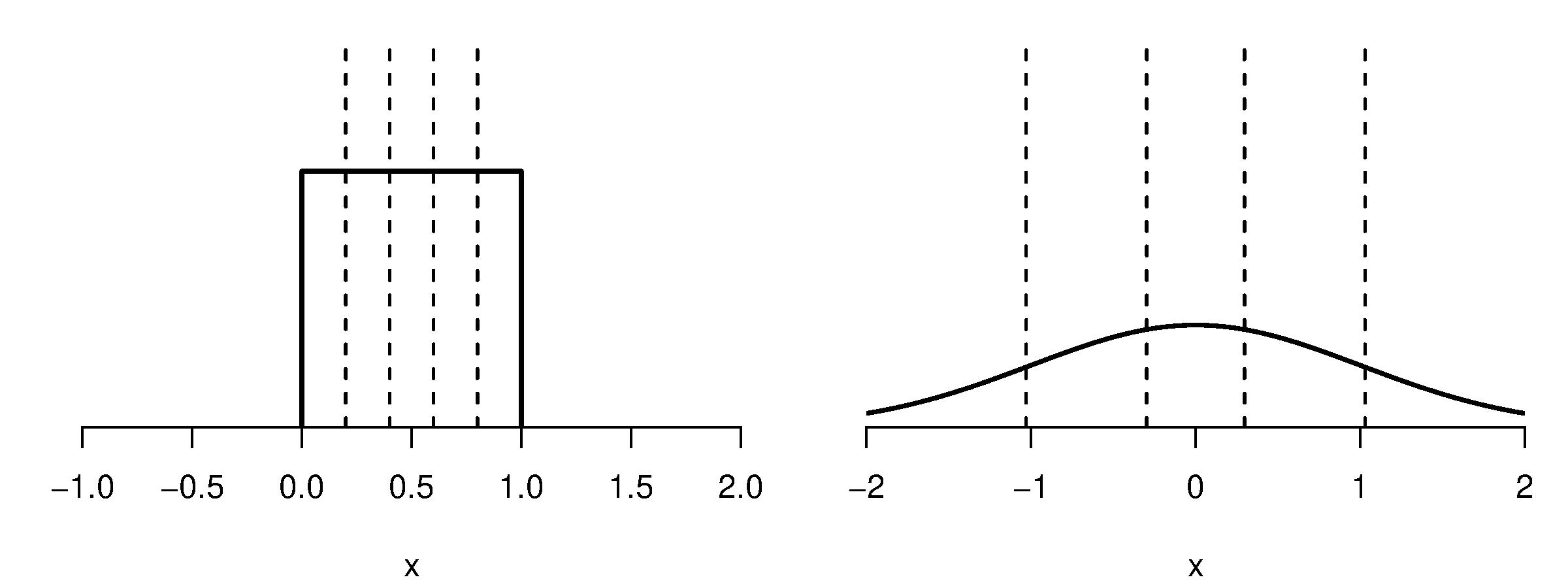}
\end{center}
\caption{The four expected order statistics of $\cU(0,1)$ (left plot) and $\cN(0,1)$ (right plot). These are equally spaced for $\cU(0,1)$ while the inner pair of expected order statistics is closer than the outer pairs of expected order statistics for $\cN(0,1)$.}
\label{fig:expected_OS_uniform_Gaussian}
\end{figure}

Additional insights come from another view of why the L-moments are centered at the uniform distributions. Note that the third and fourth theoretical L-moments are
\begin{align*}
\lambda_3&=\frac{1}{3}\left\{E\left(X_{3:3}-X_{2:3}\right)-E\left(X_{2:3}-X_{1:3}\right)
\right\},\\
\lambda_4&=\frac{1}{4}\left\{E\left(X_{4:4}-X_{3:4}\right)
-2E\left(X_{3:4}-X_{2:4}\right)+E\left(X_{2:4}-X_{1:4}\right)\right\}.
\end{align*}
These expressions indicate that if $F$ has equally spaced expected order statistics, then its third and higher order L-moments are zero. Figure \ref{fig:expected_OS_uniform_Gaussian} shows the four expected order statistics of the distributions $\cU(0,1)$ and $\cN(0,1)$ as four vertical dashed lines. Note that the vertical lines of the uniform distribution in the left plot are equally spaced; if $F\sim\cU(0,1)$, then $EX_{j:4}=j/5$ for $j=1,2,3,4$. However, for the standard Gaussian distribution in the right plot, the space between the inner pair of expected order statistics is smaller than the spaces between the two outer pairs.

This motivates us to rescale the spaces between adjacent expected order statistics by the corresponding spaces of the standard Gaussian distribution. The following theorem shows that an equivalent definition of the L-moments can be derived by re-expressing Equation (\ref{eqn:L_moments_defn_1}) in terms of spaces between expected order statistics.
\begin{thm}\label{thm:L_moments_expected_spacings}
The $r$-th theoretical L-moment $\lambda_r$ can be expressed as
\[\lambda_r=\frac{1}{r}\sum_{k=0}^{r-2}(-1)^k\left(\!
\begin{array}{c}r-2\\k\end{array}\!\right)E\left(X_{(r-k):r}-X_{(r-k-1):r}\right)\]
for $r=2,3,\cdots$.
\end{thm}
\begin{pf}
See Section \ref{supp:subsec:proof_L_moments_expected_spacings} of the Supplementary Material.\hfill\qedsymbol
\end{pf}

We first show that when any symmetric distribution is used for rescaling, the resulting measures of scale, skewness and kurtosis are linear functions of the corresponding measures based on the classical L-moments. This implies that subtracting the L-kurtosis value of a target distribution from the L-kurtosis itself is actually equivalent to rescaling based on the spaces between adjacent expected order statistics of that target distribution. Let $\delta_{i,j:k}(F)=E\left(X_{j:k}-X_{i:k}\right)$ for $1\leq i<j\leq k$ be the space between the $i$-th and $j$-th expected order statistics. Then we define the theoretical \textit{Rescaled L-moments based on the distribution $F_0$} as
\begin{equation}\label{eqn:RL_moments_based_on_F0}
\rho_{F_0,r}=\frac{1}{r}\sum_{k=0}^{r-2}\frac{(-1)^k}{\delta_{(r-k-1),(r-k):r}\left(F_0\right)}\left(\!
\begin{array}{c}r-2\\k\end{array}\!\right)E\left(X_{(r-k):r}-X_{(r-k-1):r}\right)
\end{equation}
for $r=2,3,\cdots$. We let $\rho_{F_0,1}=\lambda_1$. The corresponding theoretical \textit{Rescaled L-moment ratios based on the distribution $F_0$} are defined as $\rho_{F_0,r}^*=\rho_{F_0,r}/\rho_{F_0,2}$ for $r=3,4,\cdots$.
\begin{thm}\label{thm:Oja_criteria_RL}
Suppose that $F_0$ is a symmetric distribution. Then we have
\begin{align}\label{eqn:relationship_RL_symmetric_L}
\rho_{F_0,1}&=\lambda_1,\notag\\
\rho_{F_0,2}&=\frac{1}{\delta_{1,2:2}\left(F_0\right)}\lambda_2,\notag\\
\rho_{F_0,3}^*&=\frac{\delta_{1,2:2}\left(F_0\right)}{\delta_{1,2:3}\left(F_0\right)}\lambda_3^*,\notag\\
\rho_{F_0,4}^*&=\frac{\delta_{1,2:2}\left(F_0\right)}{5}\left\{\frac{3}{\delta_{2,3:4}\left(F_0\right)}+\frac{2}{\delta_{3,4:4}\left(F_0\right)}\right\}\lambda_4^*\notag\\
&\quad-\frac{3\delta_{1,2:2}\left(F_0\right)}{5}\left\{\frac{1}{\delta_{2,3:4}\left(F_0\right)}-\frac{1}{\delta_{3,4:4}\left(F_0\right)}\right\}.
\end{align}
These four measures satisfy Oja's criteria for a measure of location, scale, skewness and kurtosis, respectively.
\end{thm}
\begin{pf}
See Subsection \ref{supp:subsec:proof_Oja_criteria_RL} of the Supplementary Material.\hfill\qedsymbol
\end{pf}
\noindent Based on this theorem, we define the $r$-th \textit{Rescaled L-moment} (\textit{RL-moment}) as $\rho_r=\rho_{\Phi,r}$. Note that we have $\rho_4^*(F)\approx1.7560\left(\lambda_4^*(F)-\lambda_4^*(\Phi)\right)$.

Using Equation (\ref{eqn:RL_moments_based_on_F0}), it can easily be shown that the RL-moments are L-functionals such that
\[\rho_r=\int_{-\infty}^\infty\!xf(x)R_{r-1}(F(x))\,\mathrm{d}x=\int_0^1\!F^{-1}(u)R_{r-1}(u)\,\mathrm{d}u\]
where $R_r:(0,1)\rightarrow\mR$ is an $r$-th degree polynomial that satisfies
\begin{equation}\label{eqn:relationship_R_P}
R_r(u)=\sum_{k=1}^r\alpha_{k,r}P_k^*(u)
\end{equation}
for all $0<u<1$ with constants $\alpha_{j,r}\in\mR$ for $j=1,2,\cdots,r$. For example, we have $\alpha_{1,2}=0$ and $\alpha_{2,2}=1/\delta_{1,2:2}\left(\Phi\right)$. Equation (\ref{eqn:RL_moments_based_on_F0}) also implies that the RL-moments are centered at the Gaussian distributions. The first four polynomials of $R_r$ can be obtained using the results in \citet{Hosking1986} as
\[\begin{array}{ll}
R_0(u)=P_0^*(u),&R_1(u)=c_1P_1^*(u),\\
R_2(u)=c_2P_2^*(u),&R_3(u)=c_4\left\{(6c_3+2)u^3-3(3c_3+1)u^2+(3c_3+3)u-1\right\}.
\end{array}\]
where $c_1\approx0.8862,\,c_2\approx1.1816,\,c_3\approx3.4658$ and $c_4\approx1.3654$.

The RL-moment ratios based on a symmetric distribution $F_0$ of higher orders than 4 are not necessarily linear functions of the classical L-moment ratios of the same orders. For example, we have
\[\rho_{F_0,5}^*=\alpha_{F_0,5,5}^*\lambda_5^*+\alpha_{F_0,3,5}^*\lambda_3^*\]
where
\begin{align*}
\alpha_{F_0,5,5}^*&=\frac{\delta_{1,2:2}\left(F_0\right)}{7}\left\{\frac{6}{\delta_{3,4:5}\left(F_0\right)}+\frac{1}{\delta_{4,5:5}\left(F_0\right)}\right\},\\
\alpha_{F_0,3,5}^*&=-\frac{6\delta_{1,2:2}\left(F_0\right)}{7}\left\{\frac{1}{\delta_{3,4:5}\left(F_0\right)}-\frac{1}{\delta_{4,5:5}\left(F_0\right)}\right\},
\end{align*}
whose proof follows the same steps of derivation as the proof of Equation (\ref{eqn:relationship_RL_symmetric_L}). This implies that if a distribution $F$ satisfies $\lambda_3^*(F)\neq0$, then we have
\[\rho_5^*(F)=\alpha_{\Phi,5}\lambda_5^*(F)+\beta_{\Phi,5}\lambda_3^*(F)\neq\alpha_{\Phi,5}\lambda_5^*(F)=\alpha_{\Phi,5}\left\{\lambda_5^*(F)-\lambda_5^*(\Phi)\right\}\]
where the last equality results from $\lambda_5^*(\Phi)=0$. In conclusion, the RL-moment ratios based on a symmetric distribution $F_0$ are generally different from the classical L-moment ratios subtracted by their values evaluated at $F_0$.

As we mentioned in Section \ref{sec:introduction}, we are looking for moments that have zero values at the Gaussian distributions in a theoretical sense. The odd terms of the classical L-moments already have zero values at the Gaussian distributions, and the even terms can be centered at the Gaussians distributions by subtracting their numerical values at the Gaussians. However, this does not result from any characteristic of the Gaussian distributions except that those are symmetric. The RL-moments are much more intuitively appealing because of much better correspondence with the expected order statistics. Furthermore, this more mathematically principled approach gives a natural extension for developing analogs targeting any other symmetric distribution as shown in Theorem \ref{thm:Oja_criteria_RL}. Even though the RL-kurtosis turns out to be equivalent to the classical L-kurtosis, we value theoretical soundness of the RL-moments compared to numerical manipulation of the classical L-moments.

\section{TCGA data analysis}\label{sec:TCGA_data}

The goal of our TCGA data analysis is to discover biologically meaningful genes out of 16,615 genes based on their expression profiles measured on 817 breast cancer patients. As seen in Figure \ref{fig:mdp_bw_skew_L_skew_bottom}, one important biological feature that can be inferred from marginal distributions of expression profiles is the distributions of breast cancer subtypes. While cancer subtypes are of interest, there are many other biological features of interest as well, so the focus here is on distributional aspects, not simply discriminating cancer subtypes. To this end, we focus on departure from Gaussianity as a measure of screening genes in our TCGA data. As mentioned in Section \ref{sec:introduction}, we choose summary statistics to generate ranked lists of genes.

Note that skewness and bimodality of a distribution are not totally independent concepts. It was shown in \citet{Wilkins1944} and \citet{Hosking1990} that the conventional sample moments and theoretical L-moments satisfy the following relationships
\begin{align*}
\hat{\gamma}_{n,1}^2+1&\leq\hat{\gamma}_{n,2},\\
\frac{1}{4}\left(5\left(\lambda_3^*\right)^2-1\right)&\leq\lambda_4^*<1.
\end{align*}
This implies that a skewness or kurtosis estimate alone is not enough to describe skewness and bimodality of a distribution. Bimodal distributions should have low kurtosis measures when they are symmetric, but this does not hold for asymmetric bimodal distributions. Hence, we check both of the ranked lists of genes generated by skewness and kurtosis estimators and comprehensively diagnose the performances of different estimators. Joint use of skewness and kurtosis estimators such as the Jarque-Bera statistic \citep{Jarque1980} can be considered as a method of sorting genes. However, this can hinder us from interpreting the sorted lists of genes. A gene with a high Jarque-Bera statistic value can have a high degree of departure from Gaussianity, but we cannot decide whether that departure mainly comes from skewness, kurtosis or a combination of both.

\subsection{Marginal distribution plots}\label{subsec:mdp}

We investigate whether L-statistics based skewness and kurtosis estimators capture more interesting and genetically useful departures from the Gaussian distributions. Comparison of the kurtosis estimators based on their abilities in detecting bimodality did not show much visual difference, so we focus on skewness estimators here and present the results of visual comparisons between kurtosis estimators in Section \ref{supp:sec:TCGA_supplement} of the Supplementary Material. The sorted list of genes generated by the L- and RL-moments should be the same based on Theorem \ref{thm:Oja_criteria_RL}, so we omit marginal distribution analysis of the RL-moments.

On top of the conventional and Gaussian Centered L-moments, we also consider some quantile-based measures which are known to be robust as baseline measures. Bowley's skewness measure \citep{Bowley1920} is a typically used quantile-based measure of skewness defined as
\begin{equation}\label{eqn:Bowley_skewness}
\gamma_p=\frac{F^{-1}(1-p)-F^{-1}(1/2)-\left\{F^{-1}(1/2)-F^{-1}(p)\right\}}
{F^{-1}(1-p)-F^{-1}(p)}
\end{equation}
where $p$ is usually set to $0.25$, which results in $\gamma_p$ being based on quartiles.
On the other hand, Ruppert's interfractile range ratio \citep{Ruppert1987} is frequently used as a measure of kurtosis and defined as
\begin{equation}\label{eqn:Ruppert_kurtosis}
\gamma_{p_1,p_2}=\frac{F^{-1}\left(1-p_1\right)-F^{-1}\left(p_1\right)}
{F^{-1}\left(1-p_2\right)-F^{-1}\left(p_2\right)}
\end{equation}
where the parameters $p_1$ and $p_2$ were set to 0.1 and 0.3, respectively, in that paper. Note that Bowley's skewness measure is zero at the Gaussian distributions, but Ruppert's kurtosis measure is not zero there. Both Bowley's and Ruppert's estimators were shown to satisfy Oja's criteria for measures of skewness and kurtosis, respectively, in those papers.

\begin{figure}
\begin{center}
\begin{tabular}{c}
\includegraphics[width=0.96\textwidth]{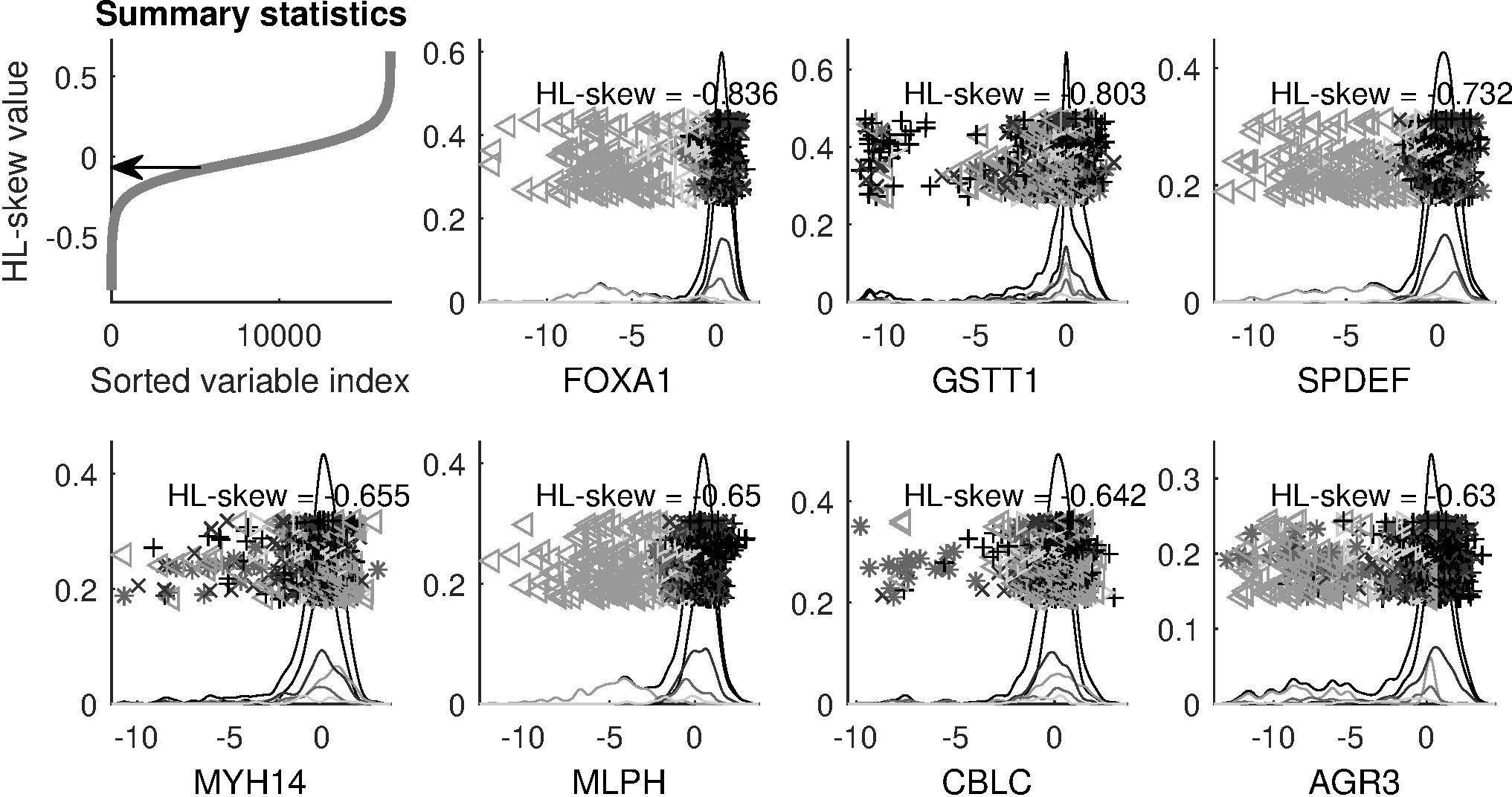}\\\\
\includegraphics[width=0.96\textwidth]{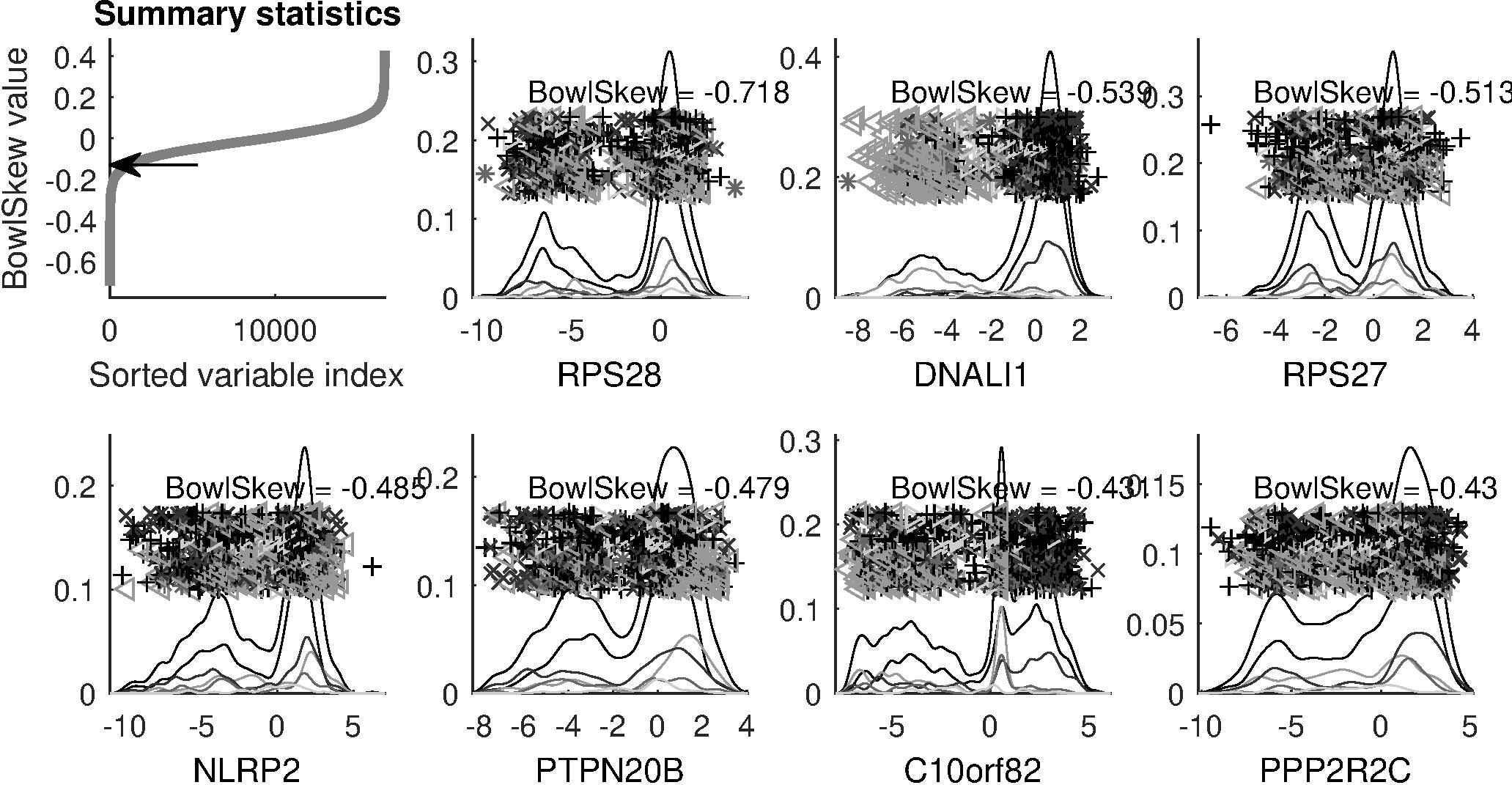}
\end{tabular}
\end{center}
\caption{The marginal distribution plots of seven genes with the smallest sample HL-skewness (top two rows) and Bowley's skewness values (bottom two rows). The seven genes in the upper two rows possess distributional skewness to the left side and focus better on subtypes, while those in the lower two rows sometimes exhibit asymmetric bimodality.}
\label{fig:mdp_bw_HL_Q_skew_bottom}
\end{figure}

In Section \ref{sec:robustness} of this paper, it will be shown that skewness measures can be ordered in terms of robustness as Bowley's, the L-, HL- and conventional skewness from the most robust to the least robust measures. To visually check whether such relationships between the influence functions is effective for finding interesting variables, we present the seven genes with the smallest HL- and Bowley's skewness in the top and bottom two rows of Figure \ref{fig:mdp_bw_HL_Q_skew_bottom}, respectively. Similarly to the genes given in the lower two rows of Figure \ref{fig:mdp_bw_skew_L_skew_bottom} screened by the L-skewness, the HL-skewness finds the genes with left-skewness in their distributional bodies and show good separation of subtypes. All the seven genes selected by the HL-skewness except SLC44A4 and GATA3 were also selected by the L-skewness. Interestingly, the three genes with the smallest L- and HL-skewness values were the same, indicating that both of the L-statistics based estimators share a similar level of robustness. However, the two genes that were selected by one estimator but not by the other motivate us to consider a quantitative analysis of the whole ranked lists generated by those two measures. A much more quantitative comparison of gene rankings is done in the next subsection.

The seven genes with the smallest Bowley's skewness estimator presented at the bottom two rows of Figure \ref{fig:mdp_bw_HL_Q_skew_bottom} show quite different patterns. Those genes generally show strong bimodality rather than distributional skewness. Especially, the genes such as RPS27 and C10orf82 may not be viewed as strongly skewed. This seems to happen because Bowley's measure ignores the distribution of data outside of the quartiles, which can result in some unintuitive notions of skewness. Whether this property is advantageous or not in screening biologically interesting genes is analyzed in the next subsection.

\subsection{Gene Set Enrichment Analysis}\label{subsec:GSEA}

One of the main goals of the TCGA data analysis done here is to assess the performances of various measures of skewness and kurtosis by their abilities to screen biologically meaningful genes. The marginal distribution plot analysis results given in the preceding subsection show that skewness estimators are able to screen genes that are related to cancer subtypes which are a type of biologically meaningful feature. We conjecture that genes with non-Gaussian distributional shape can be related to other biological features as well as cancer subtypes, and confirm this based on \textit{Gene Set Enrichment Analysis} (\textit{GSEA}) which was introduced in \citet{Subramanian2005}.

A basic goal of GSEA is to assess goodness of a ranked list of genes in terms of how well the list places biologically interesting genes near its top or bottom, i.e. how well the list finds interesting genes. Since a measure of skewness or kurtosis provides an ordering of the genes, its screening ability can be assessed by applying GSEA to the ranked list generated by it. In our GSEA, we will study performances of various measures over many gene sets published by the Broad Institute. A collection of 15,312 such gene sets, with a minimum of 15 and maximum of 10,000 genes, is available in the public database MSigDB v.6.0. Our main results are GSEA output which is how many times those gene sets are located at the top and bottom of a ranked list of genes, i.e. how many times interesting gene sets are screened by the ranked list of genes.

\begin{figure}
\begin{center}
\includegraphics[width=0.6\textwidth]{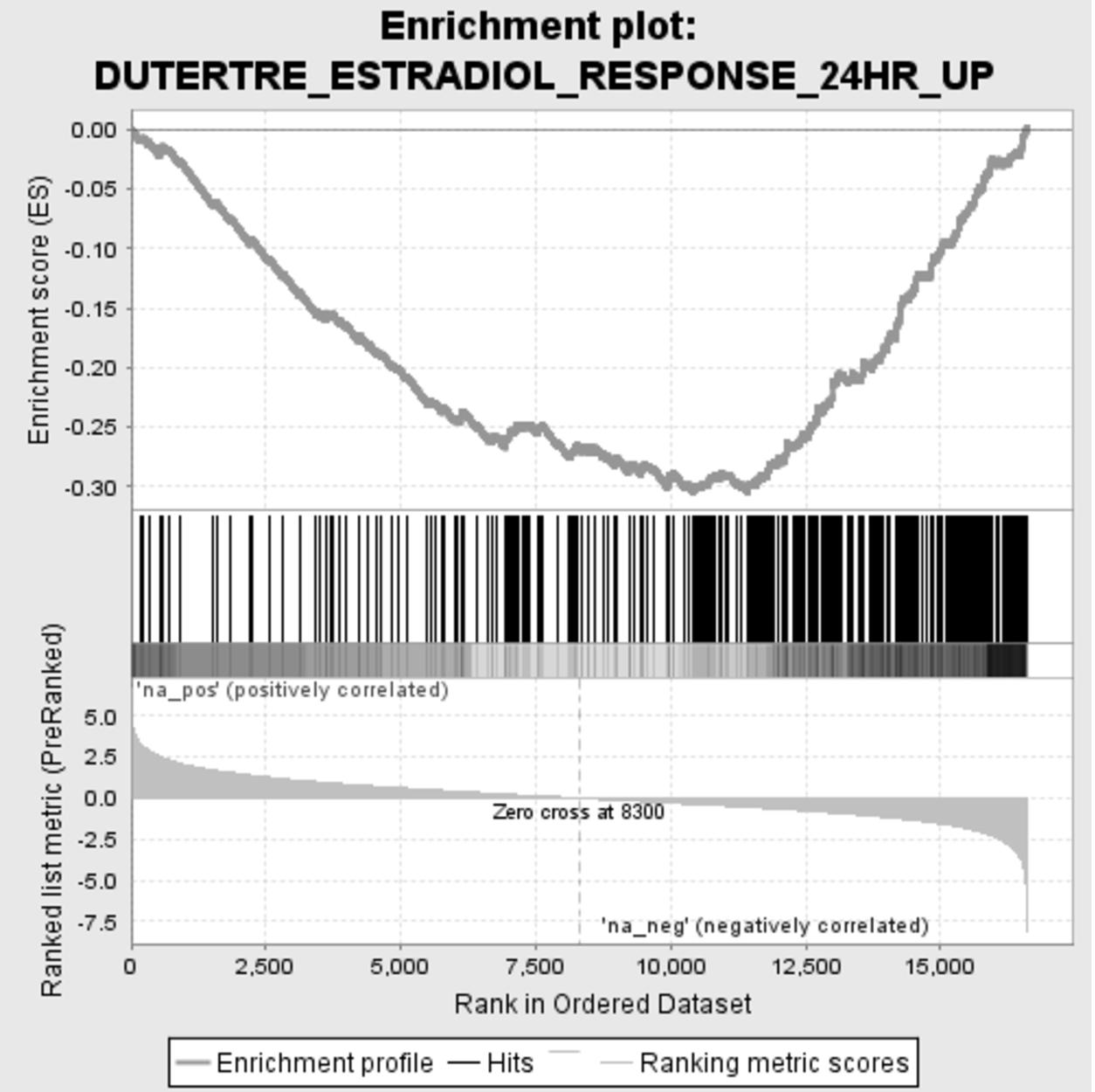}
\end{center}
\caption{The enrichment score plot. The top panel shows the values of $\text{ES}(S,n)$ and the middle panel shows the locations of interesting genes in this gene set. The bottom panel shows the sample quantiles of the HL-kurtosis estimates. It can be seen that the biologically meaningful genes mostly gather at the bottom of the ranked list yielding a negative enrichment score.}
\label{fig:es_plot}
\end{figure}

Figure \ref{fig:es_plot} shows an example, based on our data, generated by the Gene Set Enrichment software v2.2.4 released by the Broad Institute. Here we used the ranked list of genes generated by the HL-kurtosis and the independent gene set whose name is given near the top of Figure \ref{fig:es_plot}. The reason for selecting this gene set is its particularly strong result. The middle panel shows a ranked list of genes $L=\left\{g_1,g_2,\cdots,g_N\right\}$ where the locations of genes in the independent gene set $S=\left\{g_1',g_2',\cdots,g_{N'}'\right\}$ are denoted by black vertical lines. The gene set consists of 316 genes; hence there are 316 black lines. The bottom panel shows the sample quantile plot of rank metrics which are the sample HL-kurtosis values. The values are sorted in descending order and represented by the heights of gray vertical lines, yielding two half mounds of gray colors. We compute the probabilities of finding and missing interesting genes $S$ in the top $n$ list of genes extracted from the ranked list $L$ as
\begin{equation}\label{eqn:GSEA_hit_miss}
P_{\text{hit}}(S,n)=\frac{\sum_{j\leq n,g_j\in S}\left|r_j\right|^p}
{\sum_{g_j\in S}\left|r_j\right|^p},\quad
P_{\text{miss}}(S,n)=\frac{\sum_{j\leq n,g_j\notin S}1}{\sum_{g_j\notin S}1}
\end{equation}
where the parameter $p$ was recommended to take the value of one (or sometimes zero) in \citet{Subramanian2005}. As we increase the size $n$ of the top-$n$ list, both $P_{\text{hit}}(S,n)$ and $P_{\text{miss}}(S,n)$ become functions of $n$ whose maximum difference is of interest. The \textit{enrichment score} (\textit{ES}) of the gene set $S$ is defined as $\text{ES}(S)=\text{ES}(S,n')$ where $n'=\argmax_{1\leq n\leq N}\left|\text{ES}(S,n)\right|$ and
\[\text{ES}(S,n)=P_{\text{hit}}(S,n)-P_{\text{miss}}(S,n)\]
to assess the significance of difference between them. The top panel of Figure \ref{fig:es_plot} plots the values of $\text{ES}(S,n)$ as a function of $n$ so that the $\text{ES}(S)$ is obtained as the minimum value of that function. If interesting genes are gathered at the top of the ranked list, the enrichment score will be highly positive, and if they are gathered at the bottom, the score will be highly negative. It can be seen in Figure \ref{fig:es_plot} that most of the genes in that gene set have negatively large HL-kurtosis estimates, yielding the ES value of -0.304.

One important difference between our analysis and the conventional GSEA is that different measures of skewness and kurtosis have different scales. For example, the conventional kurtosis $\gamma_2$ is only bounded below by 0, but the L-kurtosis was shown in \citet{Hosking1990} to be bounded by -1 and 1. This type of difference of scales can yield unexpected results in the GSEA analysis. The paper \citet{Subramanian2005} suggested that the option $p=0$, called the \textit{classic scoring scheme} will not be much affected by the type of statistic in Equation (\ref{eqn:GSEA_hit_miss}), so we use that here.

To evaluate the statistical significance on enrichment score, $\text{ES}(S)$, we randomly permute the ranks of genes in the ranked list to compute the $k$-th permuted enrichment score $\text{ES}_k(S)$ for the $k$-th permuted list. We repeat the permutation for $K=1,000$ times to obtain the null distribution of $\text{ES}_k(S)$ and a nominal p-value of $\text{ES}(S)$. For multiple gene sets $S_m$ for $m=1,2,\cdots,M$, we control the \textit{False Discovery Rate} (\textit{FDR}) which is the proportion of permuted gene lists with larger absolute values and the same sign of enrichment scores with $\text{ES}(S)$,
\begin{align*}
\text{FDR}(S)&= \frac{\sum_{m=1}^M\sum_{k=1}^KI\left(\text{ES}_k\left(S_m\right)>\text{ES}(S)\right)}
{\sum_{m=1}^M\sum_{k=1}^KI\left(\text{ES}_k\left(S_m\right)>0\right)}\quad\text{if ES}(S)>0,\\
\text{FDR}(S)&= \frac{\sum_{m=1}^M\sum_{k=1}^KI\left(\text{ES}_k\left(S_m\right)\leq\text{ES}(S)\right)}
{\sum_{m=1}^M\sum_{k=1}^KI\left(\text{ES}_k\left(S_m\right)\leq0\right)}\quad\text{if ES}(S)\leq0.
\end{align*}
If the $\text{FDR}(S)$ is less than a given level, then it is said that the corresponding gene set $S$ is significantly enriched. We investigate both the FDR levels 0.05 and 0.25 based on the recommendations of the GSEA User Guide (\url{http://software.broadinstitute.org/gsea/doc/GSEAUserGuideFrame.html}).

\subsection{Comparison among skewness estimators}\label{subsec:GSEA_skewness}

Table \ref{tab:GSEA_results_skew_FDR025} shows the results of comparison between different skewness estimators when the FDR level is fixed at 0.25. The title row shows the names of the skewness estimators and the numbers of respective enriched gene sets. For example, the ranked list of genes generated by the HL-skewness screened 6,316 biologically meaningful gene sets, i.e. the FDR's of those gene sets were less than 0.25. The columns show, from left to right, the best to worst performing skewness estimators by the number of interesting gene sets screened by them. The title column shows the skewness estimators in the same order with the last estimator removed. We do a pairwise comparison of the performances of estimators in terms of how many gene sets are flagged as enriched and assess statistical significance using Fisher's exact test. The values given in the middle of the table are the p-values. For example, the p-value 0.6093 in the second row and the third column shows that the difference in the numbers of enriched gene sets by the HL-skewness (6,316) and L-skewness (6,271) are not significantly different. Since each pair of skewness estimators need only one comparison, we colored cells corresponding to repetitive comparisons gray. The boldfaced numbers indicate that corresponding p-values are significant, before adjusting for multiple comparisons, and will continue to be after Bonferroni correction.

It can be seen in Table \ref{tab:GSEA_results_skew_FDR025} that the HL-skewness performs the best while Bowley's skewness performs the worst. The L-statistics based skewness estimators perform better than the conventional skewness, but the degree of superiority is not statistically significant. The conventional skewness is in third place, but its inferiority to the HL-skewness yields a p-value of 0.1011, which is not significant. This motivates further investigation into relationships between those estimators, which will be shown in Table \ref{tab:GSEA_results_skew_FDR005} with the FDR level 0.05. Interestingly, Bowley's skewness is not good in terms of its screening ability. It performs the worst and its inferiority to the other estimators is very significant, as shown in the last column of the table. It seems that this results from the property of Bowley's skewness that it focuses on the locations of the sample quartiles and ignores too much information in other parts of a distribution. The TCGA data have 5 subtypes, and consideration of only the first and third sample quartiles seems to result in a strong loss of subtype information. This also implies that visual comparisons as shown in Subsection \ref{subsec:mdp}, where Bowley's skewness appeared to screen subtype driven genes quite well, are not enough to fully evaluate the relative performances between estimators.

\begin{table}
\begin{center}
\begin{tabular}{c||c|c|c|c}\hline
Estimator&HL-skewness&L-skewness&Skewness&Bowley's skewness\\
(No. of enriched genes)&(6,316)&(6,271)&(6,174)&(4,242)\\\hline\hline
HL-skewness (6,316)&\cellcolor{lightgray}&0.6093&0.1011&$\boldsymbol{<0.001}$\\\hline
L-skewness (6,271)&\cellcolor{lightgray}&\cellcolor{lightgray}&0.264&$\boldsymbol{<0.001}$\\\hline
Skewness (6,174)&\cellcolor{lightgray}&\cellcolor{lightgray}&\cellcolor{lightgray}&$\boldsymbol{<0.001}$\\\hline
\end{tabular}
\end{center}
\caption{The numbers of gene sets enriched by different skewness estimators with the FDR level 0.25 and the significance of their differences based on Fisher's exact test. The HL- and L-skewness perform the best and Bowley's skewness performs the worst. The superiority of the HL-skewness over the conventional skewness is not statistically significant, which motivates further comparison. The inferiority of Bowley's skewness to the other estimators is very significant.}
\label{tab:GSEA_results_skew_FDR025}
\end{table}

Table \ref{tab:GSEA_results_skew_FDR005} shows a similar result as Table \ref{tab:GSEA_results_skew_FDR025} at FDR level 0.05. Overall, L-statistics based estimators better capture biologically meaningful departures from Gaussianity in the direction of skewness. The L-skewness performs significantly better than all the other estimators as shown in Table \ref{tab:GSEA_results_skew_FDR005} row two. The superiority of the L-skewness over the HL-skewness is not really significant after considering multiplicity; the p-value is about 0.0498 which was shown as 0.050 in the table after rounding. The HL-skewness performs better than the conventional skewness and the degree of difference is statistically significant. This coincides with our observations made in Subsection \ref{subsec:mdp} that the conventional skewness is driven by outliers too much to be able to screen subtype-driven meaningful genes. Bowley's skewness is again not successful in screening interesting genes. It's inferiority to the other estimators is again very significant as seen in the last column. Based on Tables \ref{tab:GSEA_results_skew_FDR025} and \ref{tab:GSEA_results_skew_FDR005}, we claim that the L-statistics based skewness estimators screen interesting genes significantly better than the other estimators, but any difference between the HL- and L-skewness is hard to conclude.

\begin{table}
\begin{center}
\begin{tabular}{c||c|c|c|c}\hline
Estimator&L-skewness&HL-skewness&Skewness&Bowley's skewness\\
(No. of enriched genes)&(3,300)&(3,159)&(2,993)&(2,005)\\\hline\hline
L-skewness (3,300)&\cellcolor{lightgray}&0.050&$\boldsymbol{<0.001}$&$\boldsymbol{<0.001}$\\\hline
HL-skewness (3,159)&\cellcolor{lightgray}&\cellcolor{lightgray}&0.019&$\boldsymbol{<0.001}$\\\hline
Skewness (2,993)&\cellcolor{lightgray}&\cellcolor{lightgray}&\cellcolor{lightgray}&$\boldsymbol{<0.001}$\\\hline
\end{tabular}
\end{center}
\caption{The numbers of gene sets enriched by different skewness estimators with FDR level 0.05 and the significance of their differences based on Fisher's exact test. All differences are statistically significant. Similarly to Table \ref{tab:GSEA_results_skew_FDR025}, the L-statistics based estimators perform the best and Bowley's skewness performs the worst. This confirms that L-statistics are more useful in finding biologically meaningful genes than other estimators.}
\label{tab:GSEA_results_skew_FDR005}
\end{table}

\subsection{Comparison among kurtosis estimators}\label{subsec:GSEA_kurtosis}

In this subsection, we compare performances of kurtosis estimators in finding biologically meaningful genes in terms of GSEA. This case is more interesting than skewness, since different kurtosis measures usually have different distributional centers as mentioned in Section \ref{sec:introduction}. For example, the distributional center of the L-kurtosis is the uniform distributions, and that of Ruppert's kurtosis depends on its parameters $p_1$ and $p_2$. It was shown in Supplementary Material Section \ref{supp:sec:TCGA_supplement} that kurtosis estimators did not show much difference in the marginal distributions of the genes with the smallest kurtosis estimates. Here we check if the weak difference still holds in the quantitative analysis done by the GSEA.

Table \ref{tab:GSEA_results_kurt_FDR025} shows the comparison results when the FDR level is fixed at 0.25. The HL-kurtosis dominates all the other estimators with p-values close to zero. Ruppert's kurtosis performs significantly worse than the other measures, which confirms our observation made in Subsection \ref{subsec:GSEA_skewness} that Bowley's skewness, which is based on the sample quantiles, is not efficient in screening biologically meaningful genes. Interestingly, the L-kurtosis performs worse than the conventional kurtosis with p-value less than 0.001 (third row, fourth column), but it will be seen in Table \ref{tab:GSEA_results_kurt_FDR005} that their difference becomes insignificant as we change the FDR level to 0.05.

\begin{table}
\begin{center}
\begin{tabular}{c||c|c|c|c}\hline
Estimator&HL-kurtosis&Kurtosis&L-kurtosis&Ruppert's kurtosis\\
(No. of enriched genes)&(4,992)&(4,085)&(3,369)&(2,692)\\\hline\hline
HL-kurtosis (4,992)&\cellcolor{lightgray}&$\boldsymbol{<0.001}$&$\boldsymbol{<0.001}$&$\boldsymbol{<0.001}$\\\hline
Kurtosis (4,085)&\cellcolor{lightgray}&\cellcolor{lightgray}&$\boldsymbol{<0.001}$&$\boldsymbol{<0.001}$\\\hline
L-kurtosis (3,369)&\cellcolor{lightgray}&\cellcolor{lightgray}&\cellcolor{lightgray}&$\boldsymbol{<0.001}$\\\hline
\end{tabular}
\end{center}
\caption{The numbers of gene sets enriched by different kurtosis estimators with FDR level 0.25 and the significance of their differences based on Fisher's exact test. The HL-kurtosis performs the best and Ruppert's kurtosis performs the worst. The L-kurtosis performs worse than the conventional kurtosis (third row, fourth column).}
\label{tab:GSEA_results_kurt_FDR025}
\end{table}

Table \ref{tab:GSEA_results_kurt_FDR005} shows that the order of kurtosis estimators of their GSEA performances remains the same even though we change the FDR level to 0.05. The HL-kurtosis performs the best and Ruppert's kurtosis performs the worst at this FDR level. The conventional kurtosis again performs better than the L-kurtosis, but the degree of superiority is not statistically significant (p-value $\approx$ 0.394). This implies that the competition between the conventional kurtosis and L-kurtosis is inconclusive. Ruppert's kurtosis again performs the worst with p-values less than 0.001. Overall, it seems that the HL-kurtosis is particularly useful in screening biologically meaningful genes in the direction of kurtosis.

\begin{table}
\begin{center}
\begin{tabular}{c||c|c|c|c}\hline
Estimator&HL-kurtosis&Kurtosis&L-kurtosis&Ruppert's kurtosis\\
(No. of enriched genes)&(2,067)&(1,500)&(1,455)&(684)\\\hline\hline
HL-kurtosis (2,067)&\cellcolor{lightgray}&$\boldsymbol{<0.001}$&$\boldsymbol{<0.001}$&$\boldsymbol{<0.001}$\\\hline
Kurtosis (1,500)&\cellcolor{lightgray}&\cellcolor{lightgray}&0.394&$\boldsymbol{<0.001}$\\\hline
L-kurtosis (1,455)&\cellcolor{lightgray}&\cellcolor{lightgray}&\cellcolor{lightgray}&$\boldsymbol{<0.001}$\\\hline
\end{tabular}
\end{center}
\caption{The numbers of enriched gene sets based on classic enrichment scores. The HL-kurtosis and L-kurtosis perform the best and Ruppert's kurtosis performs the worst. The superiority of the HL-kurtosis over the conventional kurtosis is statistically significant, which indicates that L-statistics perform better than sample moments or sample quantiles in screening biologically meaningful genes.}
\label{tab:GSEA_results_kurt_FDR005}
\end{table}

\section{Estimation of the Gaussian Centered L-moments}\label{sec:estimation_GCL_moments}

This section contains details of the derivation of the methods used in this paper. One of the main strengths of the Gaussian Centered L-moments is their interpretability; their values indicate direction and magnitude of departure from the Gaussian distributions. For such a strength to be effective in real data analysis, the sample Gaussian Centered L-moments should converge to theoretical parallels yielding the desired interpretability. We first focus on the HL-moments. There are multiple ways to estimate L-functionals in Equation (\ref{eqn:L_functional}) by L-statistics in Equation (\ref{eqn:L_statistic}). Our estimator starts from the following approximation
\begin{align}\label{eqn:sample_HL_approximation}
\eta_r=E_F\left(H_{r-1}\left(\Phi^{-1}(F(X))\right)X\right)
&\approx\frac{1}{n}\sum_{i=1}^nH_{r-1}\left(\Phi^{-1}\left(F\left(X_i\right)\right)\right)X_i\notag\\
&=\frac{1}{n}\sum_{i=1}^nH_{r-1}\left(\Phi^{-1}\left(F\left(X_{i:n}\right)\right)\right)X_{i:n}
\end{align}
where the approximation can be replaced by the almost sure convergence as $n\rightarrow\infty$ when suitable assumptions are made on the distribution $F$. Note that the Brown-Hettmansperger estimator in Equation (\ref{eqn:sample_HL_moments_BH}) results from a different approximation
\[E_F\left(H_{r-1}\left(\Phi^{-1}(F(X))\right)X\right)\approx E_{F_n}\left(H_{r-1}\left(\Phi^{-1}(F(X))\right)X\right)\]
where $F_n$ is the EDF. For the last expression in Equation (\ref{eqn:sample_HL_approximation}) to actually play the role of an estimator, the terms including $F$ should be estimated. A typical L-statistic can be derived from the following approximation
\begin{align}\label{eqn:L_statistics_HL_moments}
\frac{1}{n}\sum_{i=1}^nH_{r-1}\left(\Phi^{-1}\left(\underline{F\left(X_{i:n}\right)}\right)\right)
X_{i:n}
&\approx\frac{1}{n}\sum_{i=1}^nH_{r-1}\left(\Phi^{-1}\left(
\underline{E\left(F\left(X_{i:n}\right)\right)}\right)\right)X_{i:n}\notag\\
&=\frac{1}{n}\sum_{i=1}^nH_{r-1}\left(\Phi^{-1}\left(E\left(U_{i:n}\right)\right)\right)X_{i:n}\notag\\
&=\frac{1}{n}\sum_{i=1}^nH_{r-1}\left(\Phi^{-1}\left(\frac{i}{n+1}\right)\right)X_{i:n}
\end{align}
where the underlined expressions present approximated (left) and approximating (right) terms and $U_{i:n}$ is the $i$-th uniform order statistic. This is consistent with the idea given in \citet{Harter1984} that L-statistics base their performance on how well the coefficient $i/(n+1)$ approximates $F\left(X_{i:n}\right)$.

Another estimator can be obtained from a different approximation in Equation (\ref{eqn:L_statistics_HL_moments}) as
\begin{align}\label{eqn:sample_HL_moments_origin}
\frac{1}{n}\sum_{i=1}^n\underline{H_{r-1}\left(\Phi^{-1}\left(F\left(X_{i:n}\right)\right)\right)}
X_{i:n}&\approx\frac{1}{n}\sum_{i=1}^n
\underline{E\left(H_{r-1}\left(\Phi^{-1}\left(F\left(X_{i:n}\right)\right)\right)\right)}X_{i:n}\notag\\
&=\frac{1}{n}\sum_{i=1}^n
E\left(H_{r-1}\left(Z_{i:n}\right)\right)X_{i:n}
\end{align}
where $Z_{i:n}$ is the $i$-th standard Gaussian order statistic. The key idea is that careful choice of location of the expectation can increase accuracy of approximation of an L-functional by an L-statistic. Since the quantile function $\Phi^{-1}$ is a highly nonlinear function, taking expectation outside $\Phi^{-1}$ can yield better approximation in Equation (\ref{eqn:sample_HL_moments_origin}). The \textit{sample Hermite L-moments} (\textit{sample HL-moments}) are defined as
\begin{equation}\label{eqn:sample_HL_moments}
\hat{\eta}_{n,r}=\frac{1}{n}\sum_{i=1}^nE\left(H_{r-1}\left(Z_{i:n}\right)\right)X_{i:n}.
\end{equation}
This can be understood as the inner product between the order statistics $X_{i:n}$ and polynomials of the expected order statistics of the standard Gaussian distribution. By changing the degree of the polynomial, $r$, different distributional aspects of $F$ are compared with the standard Gaussian distribution $\Phi$. For example, the third and fourth sample HL-moments are
\begin{align*}
\hat{\eta}_{n,3}&=\frac{1}{n}\sum_{i=1}^n
\left\{E\left(Z_{i:n}^2\right)-1\right\}X_{i:n},\\
\hat{\eta}_{n,4}&=\frac{1}{n}\sum_{i=1}^n
\left\{E\left(Z_{i:n}^3\right)-3E\left(Z_{i:n}\right)\right\}X_{i:n}.
\end{align*}
We perform numerical integration to compute the values $E\left(Z_{i:n}^k\right)$ for $k=1,2,\cdots$. Note that the sample HL-moments are biased, e.g. we have $E_\Phi\left(\hat{\eta}_{20,4}\right)\approx-0.2833$ and $E_\Phi\left(\hat{\eta}_{50,4}\right)\approx-0.1733$. To correct these biases, we derive their means by Monte-Carlo simulation based on 10,000 simulations and subtracted the means from estimates. Note that \citet{Brown1996} adopted the same approach for their estimators. The comparison between the estimators in Equations (\ref{eqn:L_statistics_HL_moments}), (\ref{eqn:sample_HL_moments_origin}) and the estimator of \citet{Brown1996} given in Equation (\ref{eqn:sample_HL_moments_BH}) is given in Subsection \ref{supp:subsec:comparison_HL_GSEA} of the Supplementary Material.

Since the $r$-th RL-moment is a linear combination of the classical L-moments as shown in Subsection \ref{subsec:RL_moments}, the $r$-th \textit{sample Rescaled L-moment} (\textit{sample RL-moment}) is naturally derived as a linear combination of the classical sample L-moments. From Equation (\ref{eqn:relationship_R_P}), we define the $r$-th sample RL-moment as $\hat{\rho}_{n,1}=\hat{\lambda}_{n,1}$ and
\[\hat{\rho}_{n,r}=\sum_{k=1}^r\alpha_{k,r}\hat{\lambda}_{n,r}\]
for $r=2,3,\cdots$. As a result, the estimators of the RL-moments-based measures of location, scale, skewness and kurtosis given in Theorem \ref{thm:Oja_criteria_RL} are derived as follows,
\begin{align}\label{eqn:relationship_sample_RL_L}
\hat{\rho}_{n,1}&=\hat{\lambda}_{n,1},\notag\\
\hat{\rho}_{n,2}&=\frac{1}{\delta_{1,2:2}\left(\Phi\right)}\hat{\lambda}_{n,2},\notag\\
\hat{\rho}_{n,3}^*&=\frac{\delta_{1,2:2}\left(\Phi\right)}{\delta_{1,2:3}\left(\Phi\right)}\hat{\lambda}_{n,3}^*,\notag\\
\hat{\rho}_{n,4}^*&=\frac{\delta_{1,2:2}\left(\Phi\right)}{5}\left\{\frac{3}{\delta_{2,3:4}\left(F_0\right)}+\frac{2}{\delta_{3,4:4}\left(\Phi\right)}\right\}\hat{\lambda}_{n,4}^*
-\frac{3\delta_{1,2:2}\left(\Phi\right)}{5}\left\{\frac{1}{\delta_{2,3:4}\left(\Phi\right)}-\frac{1}{\delta_{3,4:4}\left(\Phi\right)}\right\}.
\end{align}

We have the following theorem on asymptotic Gaussianity of the sample HL-moment ratios, RL-skewness and RL-kurtosis.
\begin{thm}\label{thm:asymptotic_sample_HL_RL}
Let $r_1,r_2=3,4,\cdots$ such that $r_1\neq r_2$. If $E\left|X_1\right|^{2+\epsilon}<\infty$ for some $\epsilon>0$, then we have
\[n^{1/2}\left(\left(\begin{array}{c}\hat{\eta}_{n,r_1}^*\\\hat{\eta}_{n,r_2}^*\end{array}\right)
-\left(\begin{array}{c}\eta_{r_1}^*\\\eta_{r_2}^*\end{array}\right)\right)
\converge^d\cN\left(0,\Psi^H\right)\]
where the matrix $\Psi^H$ is defined in Subsubsection \ref{supp:subsubsec:proof_asymptot_sample_HL} of the Supplementary Material.
If $E\left|X_1\right|^2<\infty$, then we have
\[n^{1/2}\left(\left(\begin{array}{c}\hat{\rho}_{n,3}^*\\\hat{\rho}_{n,4}^*\end{array}\right)
-\left(\begin{array}{c}\rho_3^*\\\rho_4^*\end{array}\right)\right)
\converge^d\cN\left(0,\Psi^R\right).\]
The matrix $\Psi^R$ is defined in Subsubsection \ref{supp:subsubsec:proof_asymptot_sample_RL} of the Supplementary Material.
\end{thm}
\begin{pf}
See Subsection \ref{supp:subsec:proof_asymptotic_sample_HL_RL} of the Supplementary Material.\hfill\qedsymbol
\end{pf}

\section{Robustness}\label{sec:robustness}

Now we carefully study the relative robustness of these methods. We use the \textit{influence function} as a primary tool for robustness analysis. Note from \citet{Huber2009} that the influence function of a functional $\theta$ evaluated at a distribution $F$ is defined as
\begin{equation}\label{eqn:influence_functions}
\text{IF}(x;F,\theta)=\lim_{\eps\downarrow0}\frac{\theta\left(F_{\eps,x}\right)-\theta(F)}{\eps}
\end{equation}
where $F_{\eps,x}=(1-\eps)F+\eps\delta_x$ and $\delta_x$ is a degenerate distribution putting mass 1 at the point $x$. As can be seen from Equation (\ref{eqn:influence_functions}), the influence function measures the effect of contamination of a distribution by a point $x$ on the functional $\theta$. Hence, if a functional is sensitive to an outlier, its influence function should have large values for large absolute values of $x$. If the distribution $F$ changes, the same outlier $x$ can affect the functional in a different way. In this paper, we compare the robustness of various measures of skewness and kurtosis based on their influence functions evaluated at a family of distributions.

The papers \citet{Groeneveld1991} and \citet{Ruppert1987} compared various measures of skewness and kurtosis, respectively, using the influence function. As a criterion of comparison, both the papers compared the degrees of polynomials that are \textit{asymptotic tight bounds} of the influence functions. Suppose that $J_1,J_2:\mR\rightarrow\mR_+$ are two functions where $\mR_+=\{x\in\mR|x\geq0\}$. We write $J_1(x)=\Theta\left(J_2(x)\right)$ to mean that there exist $a_1,a_2>0$ and $x'>0$ such that
\[a_1J_2(x)\leq J_1(x)\leq a_2J_2(x)\]
for all $|x|\geq x'$. This roughly means that asymptotic behavior of both the functions $J_1$ and $J_2$ are the same. Using the asymptotic tight bounds, those papers compared the robustness of different measures. For example, if two functionals $\theta_1$ and $\theta_2$ satisfy $\left|\text{IF}\left(x;F,\theta_1\right)\right|=\Theta(|x|)$ and $\left|\text{IF}\left(x;F,\theta_2\right)\right|=\Theta\left(x^2\right)$, then $\theta_1$ was considered to be more robust than $\theta_2$ for the distribution $F$.

An interesting family of distributions for evaluation of the influence functions is \textit{Tukey's $g$ and $h$ distributions} which contain all the transformed random variables of the form
\[\left(\frac{e^{gZ}-1}{g}\right)\exp\left[\frac{hZ^2}{2}\right]\]
where $g\in\mR,h\geq0$ and $Z$ is the standard Gaussian random variable; see \citet{Martinez1984} for detailed explanation. By convention, the case when $g=0$ is defined as the random variable $Z\exp\left(hZ^2/2\right)$. We denote the distribution function of $\text{Tukey}(g,h)$ by $T_{g,h}$. An important special case, where the distributions are symmetric, $T_{0,h}$ is called \textit{Tukey's $h$ distributions}. As shown in \citep{Brys2004,Brys2006}, Tukey's $g$ and $h$ family is ideal for our study because it allows direct application of Oja's criteria. For example, $\Phi$ does not have more kurtosis than $T_{0,h}$ (Definition \ref{defn:Oja_criteria}.d) if we have $g=0$ and $h>0$. This implies that Tukey's $h$ distributions have heavier tails than the Gaussian distributions in Oja's sense. Note that heavier tails of a distribution indicate a higher chance of existence of extreme outliers. This indicates the value of checking the influence functions not only for the standard Gaussian distribution but also for Tukey's $h$ distributions with $h>0$.

Before we derive the influence functions of measures of skewness and kurtosis based on the Gaussian Centered L-moments, we introduce the previous results for the conventional skewness and kurtosis. Note that \citet{Ruppert1987} used the notion of \textit{symmetric influence function} defined as
\[\text{SIF}(x;F,\theta)=
\lim_{\eps\downarrow0}\frac{\theta\left(\left(F_{\eps,x}+F_{\eps,-x}\right)/2\right)-\theta(F)}{\eps}.\]
The symmetric influence function measures the sensitivity of a functional to symmetric contamination by points $-x$ and $x$ so it is suitable for comparison of kurtosis measures.
\begin{thm}[\citet{Groeneveld1991},\citet{Ruppert1987}]\label{thm:IF_conventional_skew_kurt}
Suppose that $F$ is a symmetric distribution such that $\mu(F)=0$ and $\sigma^2(F)=1$. Then we have
\[\text{IF}\left(x;F,\gamma_1\right)=x^3-3x=H_3(x),\,
\text{SIF}\left(x;F,\gamma_2\right)=x^4-6x^2+3=H_4(x).\tag*{$\blacksquare$}\]
\end{thm}
Before we derive the influence functions of the measures of skewness and kurtosis based on the L-, RL- and HL-moments at various distributions, we first show the relationships between their influence functions and symmetric influence functions.
\begin{thm}\label{thm:relation_IF_SIF_GCL_moments}
If $F$ is a symmetric distribution, we have
\begin{align*}
\text{SIF}\left(x;F,\lambda_4^*\right)&=\text{IF}\left(x;F,\lambda_4^*\right),\\
\text{SIF}\left(x;F,\rho_4^*\right)&=\text{IF}\left(x;F,\rho_4^*\right),\\
\text{SIF}\left(x;F,\eta_4^*\right)&=\text{IF}\left(x;F,\eta_4^*\right).
\end{align*}
\end{thm}
\begin{pf}
See Subsection \ref{supp:subsec:proof_relation_IF_SIF_GCL_moments} of the Supplementary Material.
\hfill\qedsymbol
\end{pf}

As noted above, we adopt Tukey's $h$ distributions as grounds for comparison between different measures of skewness and kurtosis. The main result is given as the following theorem.
\begin{thm}\label{thm:IF_Tukey_g_h}
We have
\begin{align*}
\left|\text{IF}\left(x;T_{0,h},\gamma_{0.25}\right)\right|=
\left|\text{IF}\left(x;T_{0,h},\gamma_{0.1,0.3}\right)\right|&=\Theta(1),\\
\left|\text{IF}\left(x;T_{0,h},\lambda_r^*\right)\right|=
\left|\text{IF}\left(x;T_{0,h},\rho_r^*\right)\right|&=\Theta(|x|),\\
\left|\text{IF}\left(x;T_{0,h},\eta_r^*\right)\right|&=
\Theta\!\left(|x|\left\{\log(|x|+1)\right\}^{(r-1)/2}\!\right)\\
\left|\text{IF}\left(x;T_{0,h},\gamma_1\right)\right|&=\Theta\left(|x|^3\right)\\
\left|\text{IF}\left(x;T_{0,h},\gamma_2\right)\right|&=\Theta\left(|x|^4\right)
\end{align*}
for all $h>0$ and $r=3,4,\cdots$.
\end{thm}
\begin{pf}
See Subsection \ref{supp:subsec:proof_IF_Tukey_g_h} of the Supplementary Material.\hfill\qedsymbol
\end{pf}
Theorem \ref{thm:IF_Tukey_g_h} implies that the Gaussian Centered L-moments are much more robust than the conventional skewness and kurtosis on Tukey's $h$ distributions. Note that the influence function of $\eta_r^*$, $\text{IF}\left(x;T_{0,h},\eta_r^*\right)$, does not depend on the parameter $h$. This indicates that even slightly heavier tails than the standard Gaussian distribution result in better robustness of the HL-moments than the conventional moments. The RL-moments have the same levels of robustness with the L-moments, which makes sense since the definitions of the RL-moments have the same forms as the L-moments except the constants of the polynomials inside their definitions. An interesting point is that the RL-moments are more robust than the HL-moments. This shows that even though different functionals have the same distributional centers and are estimated by L-statistics, they can have different levels of robustness.

The influence functions of Bowley's and Ruppert's measures, $\gamma_{0.25}$ and $\gamma_{0.1,0.3}$, given in Theorem \ref{thm:IF_Tukey_g_h} were obtained from the results of \citet{Ruppert1987} and \citet{Groeneveld1991}. Those papers showed that the influence functions of such quantile-based measures evaluated at symmetric distributions are bounded, while we showed in Theorem \ref{thm:IF_Tukey_g_h} of this paper and will complement in Theorem \ref{supp:thm:IF_standard_Gaussian} of the Supplementary Material that the L-statistics based measures have unbounded influence functions for some symmetric distributions. This justifies the use of quantile based measures as baseline robust estimators in the TCGA data analysis given in Section \ref{sec:TCGA_data}.

\bigskip
\begin{center}
{\large\bf SUPPLEMENTARY MATERIAL}
\end{center}

\begin{description}
\item[Supplementary Material:] All the proofs of the theorems in the manuscript. Figures of marginal distribution plots of the TCGA lobular freeze data. (pdf file)
\end{description}

\bigskip
\begin{center}
{\large\bf ACKNOWLEDGEMENT}
This research was supported in part by the National Science Foundation under Grant No. 1016441, 1512945, 1633074, DMS-1613112 and IIS-1633212. We give our special thanks to Youli Xia who provided input in GSEA data analysis.
\end{center}

\bibliographystyle{Chicago}
\bibliography{GCL_moments_arXiv}

\end{document}